\documentclass[sn-mathphys]{sn-jnl}

 \usepackage{etoolbox}

    \makeatletter
\patchcmd{\ps@headings}
{\hbox to \hsize{\hfill Springer Nature 2021 \LaTeX\ template\hfill}}
{\hbox to \hsize{}}
{}
{}
\patchcmd{\ps@headings}
{\hbox to \hsize{\hfill Springer Nature 2021 \LaTeX\ template\hfill}}
{\hbox to \hsize{}}
{}
{}
\patchcmd{\ps@titlepage}
{\hbox to \hsize{\hfill Springer Nature 2021 \LaTeX\ template\hfill}}
{\hbox to \hsize{}}
{}
{}
\makeatother

\usepackage{lineno}
\usepackage{graphicx}
\usepackage{amssymb}
\usepackage{amsmath}
\usepackage{mathrsfs}

\usepackage{CJKutf8}

\newcommand{\ket}[1]{\mbox{\ensuremath{\vert #1 \rangle}}}

\let\oldequation\equation
\let\oldendequation\endequation

\renewenvironment{equation}
  {\linenomathNonumbers\oldequation}
  {\oldendequation\endlinenomath}

\begin{document}

\title[Mediated interactions between Fermi polarons]
{Mediated interactions between Fermi polarons and the role of impurity quantum statistics}

\author[1,2]{\fnm{Cosetta} \sur{Baroni}}
\author[1,2]{\fnm{Bo} \sur{Huang} (\begin{CJK*}{UTF8}{gbsn}黄博\end{CJK*})}
\author[1,2]{\fnm{Isabella} \sur{Fritsche}}
\author[1,2]{\fnm{Erich} \sur{Dobler}}
\author[1,2]{\fnm{Gregor} \sur{Anich}}
\author[1,2]{\fnm{Emil} \sur{Kirilov}}
\author[1,2]{\fnm{Rudolf} \sur{Grimm}}
\author[3]{\fnm{Miguel A.} \sur{Bastarrachea-Magnani}}
\author[4]{\fnm{Pietro} \sur{Massignan}}
\author[5]{\fnm{Georg} \sur{Bruun}}

\affil[1]{\orgdiv{Institut f\"ur Quantenoptik und Quanteninformation (IQOQI)}, \orgname{\"Osterreichische Akademie der Wissenschaften}, \orgaddress{\street{Technikerstra{\ss}e 21a}, \city{Innsbruck}, \postcode{6020}, \country{Austria}}}

\affil[2]{\orgdiv{Institut f\"ur Experimentalphysik}, \orgname{Universit\"at Innsbruck}, \orgaddress{\street{Technikerstra{\ss}e 25}, \city{Innsbruck}, \postcode{6020}, \country{Austria}}}

\affil[3]{\orgdiv{Departamento de F\'isica}, \orgname{Universidad Aut\'onoma Metropolitana-Iztapalapa}, \orgaddress{\street{Av. Ferrocarril San Rafael Atlixco 186}, \city{Ciudad de M\'exico}, \postcode{09310}, \country{Mexico}}}

\affil[4]{\orgdiv{Departament de F\'isica}, \orgname{Universitat Polit\`ecnica de Catalunya}, \orgaddress{\street{Campus Nord B4-B5}, \city{Barcelona}, \postcode{08034}, \country{Spain}}}

\affil[5]{\orgdiv{Department of Physics and Astronomy}, \orgname{Aarhus University}, \orgaddress{\street{Ny Munkegade 120}, \city{Aarhus}, \postcode{8000}, \country{Denmark}}}

\abstract{

The notion of quasi-particles is essential for understanding the behaviour of complex many-body systems. A prototypical example of a quasi-particle, a polaron, is an impurity strongly interacting with a surrounding medium. Fermi polarons, created in a Fermi sea, provide a paradigmatic realization of this concept. As an inherent and important property such quasi-particles interact with each other via modulation of the medium. While quantum simulation experiments with ultracold atoms have significantly improved our understanding of individual polarons, the detection of their interactions has remained elusive in these systems. Here, we report the unambiguous observation of mediated interactions between Fermi polarons consisting of K impurities embedded in a Fermi sea of Li atoms. Our results confirm two landmark predictions of Landau's Fermi-liquid theory: the shift of the polaron energy due to mediated interactions, linear in the concentration of impurities, and its sign inversion with impurity quantum statistics. For weak to moderate interactions between the impurities and the medium, we find excellent agreement with the static (zero-momentum and energy) predictions of Fermi-liquid theory. For stronger impurity-medium interactions, we show that the observed behaviour at negative energies can be explained by a more refined many-body treatment including retardation and molecule formation

}

\maketitle 

The theory of Fermi liquids, introduced by L.\ Landau in the late 1950's, lies at the heart of our understanding of  quantum many-body  systems such as atomic nuclei, liquid Helium-3, electrons in solid-state materials, Kondo impurities, and neutron stars~\cite{Nozieres1964book,Baym2004lfl}.  
Landau’s key intuition was to explain the behaviour of many complex many-body systems in terms of elementary excitations dubbed \textit{quasi-particles}. The strong interactions between the fundamental constituents may be absorbed into few parameters characterising the quasi-particles, such as their energy and effective mass. This allows for a dramatically simplified description akin to one of weakly interacting particles.

An inherent property of quasi-particles is that they mutually interact by modulating the surrounding medium. For instance, phonon-mediated interactions give rise to conventional superconductivity, whereas spin waves are conjectured to mediate the interactions that lead to high-temperature superconductivity. The interaction between quasi-particles due to the exchange of particle-hole excitations in a surrounding Fermi sea is crucial for understanding both the equilibrium and dynamical properties of Fermi liquids. It is of particular importance for the emergence of collective modes in Fermi liquids ~\cite{Baym2004lfl}, the appearance of giant magneto-resistance, as well as for the coupling between nuclear magnetic moments as predicted by Ruderman-Kittel-Kasuya-Yosida (RKKY)~\cite{Nozieres1964book}.

The experimental approach of quantum simulation with ultracold atoms has substantially improved our understanding of single quasi-particles. Here, the quasi-particle is formed by an impurity atom surrounded by a Fermi gas or a Bose-Einstein condensate (BEC) \cite{Schirotzek2009oof,Kohstall2012mac,Hu2016bpi,Jorgensen2016ooa,Scazza2017rfp} and is referred to as \textit{Fermi polaron} or \textit{Bose polaron}, respectively. In a recent experiment \cite{DeSalvo2019oof}, fermion-mediated interactions between atoms in a BEC were observed under conditions of rather weak interspecies interaction. In the strongly interacting regime, interferometric spectroscopy in the time domain \cite{Cetina2016umb} revealed a loss of contrast in the case of a non-zero impurity concentration, but an unambiguous detection of the quasi-particle interaction in the strongly interacting regime has remained elusive \cite{Schirotzek2009oof,Scazza2017rfp}. 
Experiments on out-of-equilibrium two-dimensional materials demonstrated interactions between excitons mediated by electrons, but with a sign opposite to the one predicted by Fermi-liquid theory for a system in equilibrium \cite{Muir2022ibf,Tan2022bpi}. 

In this work, we report on the observation of mediated interactions between Fermi polarons formed by either bosonic $^{41}$K or fermionic $^{40}$K atoms in a Fermi sea of $^{6}$Li atoms.  
By choosing the K isotope, our system offers the unique possibility to change the impurity quantum statistics, leaving everything else essentially unchanged. From the observed linear shift of the polaron energy with increasing concentration we extract a momentum averaged quasi-particle interaction. Taking advantage of the exceptional degrees of control offered by the ultracold quantum gas mixture, we examine this interaction as a function of the sign and strength of K-Li interaction and find excellent agreement with the predictions of Fermi liquid theory for weak to moderate interaction strengths. This includes a predicted sign reversal of the mediated quasi-particle interaction with the impurity quantum statistics. For strong interactions, we observe deviations from Landau theory, which may be explained by the disappearance of the polaron and the emergence of molecular excitations on the attractive side.

Within Fermi-liquid theory, an impurity ($\downarrow$) interacting with a degenerate Fermi sea (FS) of ($\uparrow$) particles forms a quasi-particle. The energy of such a Fermi polaron depends on the strength of the interaction between the impurity and the particles in the FS, which, for a short range interaction, can be characterized by the dimensionless parameter $X = -1/(k_\text{F}a)$. 
Here, $a$ is the $s$-wave interspecies scattering length 
and $k_{\text{F}} = (6\pi^2 n_\uparrow)^{1/3} = \sqrt{2m_\uparrow \epsilon_{\text{F}}}/\hbar$ is the Fermi wave number, with $m_\uparrow$ the mass of particles forming the FS, $n_\uparrow$ their number density, $\epsilon_{\text{F}}$ the Fermi energy of the FS, and $\hbar$ the reduced Planck constant. 
The energy $\epsilon_{\mathbf k\downarrow}^0$ of a single  polaron with 
momentum ${\mathbf k}$ and mass $m_\downarrow$ is then, for a given mass ratio $m_\downarrow/m_\uparrow$, a  universal function of $X$~\cite{Chevy2006upd, Lobo2006nso, Combescot2007nso, Combescot2008nso, Prokofev2008fpp, Massignan2011rpa, Trefzger2012iia, Schmidt2018umb} which has been widely studied in previous experiments \cite{Schirotzek2009oof, Kohstall2012mac, Scazza2017rfp, Yan2019bau, Fritsche2021sab}. 

By increasing the concentration of the impurities, ${\mathcal C}=n_\downarrow/n_\uparrow$, more polarons are formed, and they interact with each other via density modulations in the medium. According to Landau's theory, the energy needed for creating a polaron with momentum ${\mathbf k}$ can, for small ${\mathcal C}$, be written as~\cite{Baym2004lfl}  
\begin{equation}
\epsilon_{\mathbf k\downarrow}=\epsilon_{\mathbf k\downarrow}^0
+
\sum_{\mathbf k'} f_{\mathbf k,\mathbf k'} n_{\mathbf k'\downarrow},
\label{eq:LandauQPenergy}
\end{equation} 
where $f_{\mathbf k,\mathbf k'}$ represents the interaction between two polarons with momenta ${\mathbf k}$ and ${\mathbf k}'$ mediated by atoms of the medium,  
and $n_{\mathbf k\downarrow}$ is the momentum-resolved polaron density.
One can calculate the quasi-particle interaction using perturbation theory in the atom-impurity coupling constant $g_{\uparrow\downarrow}=2\pi \hbar^2 a/m_r$, where $m_r=1/(m_\uparrow^{-1}+m_\downarrow^{-1})$ is the two-body reduced mass. This yields to second order~\cite{Yu2010con,Yu2012iii,Mora2010npo}
\begin{equation}
f_{\mathbf k,\mathbf k'}=\pm g_{\uparrow\downarrow}^2\,\chi(\mathbf{k}-\mathbf{k}',k^2/2m_\downarrow-{k'}^2/2m_\downarrow)
\label{Interaction}
\end{equation}
with $\chi({\mathbf p},\omega)$ the Lindhard function~\cite{Giuliani2005} and the upper and lower signs refer to bosonic and fermionic impurities, respectively (see Supplementary Information for further details). Equation \eqref{Interaction} is  the well-known RKKY interaction~\cite{Nozieres1964book}. 
Within Fermi liquid theory, equation \eqref{Interaction} is naturally generalized to dressed quasi-particles by replacing $g_{\uparrow\downarrow}$ (the coupling constant between bare impurities and atoms of the medium) with the polaron-atom interaction  $\partial  \epsilon_{\text{F}}/\partial n_\downarrow$, where
$n_\downarrow=\sum_{\mathbf k}n_{\mathbf k\downarrow}$ is the total impurity density \cite{Yu2012iii}. In the limit of vanishing energy and momenta one finds
\begin{equation}
f_{0}=\mp \frac{2\epsilon_{\text{F}}}{3n_\uparrow}(\Delta N)^2,
\label{Interaction2}
\end{equation}
where $\Delta N=-\partial \epsilon^0_{0\downarrow}/\partial \epsilon_{\text{F}}$ is the number of particles in the dressing cloud around a zero-momentum impurity~\cite{Massignan2005vra}.
Equations \eqref{Interaction}-\eqref{Interaction2} rigorously demonstrate that the sign of the induced interaction depends explicitly on the quantum statistics of the impurities.

To facilitate comparison with our spectroscopy experiment, which is not momentum-resolved,
we average equation~\eqref{eq:LandauQPenergy} over momentum and write 
\begin{equation}
\epsilon_{\downarrow}=\epsilon_{\downarrow}^0+\bar f n_\uparrow {\mathcal C},
\label{eq:QPenergyExperiment}
\end{equation} 
where $\bar f$ represents a  momentum average of the quasi-particle interaction $f_{\mathbf k,\mathbf k'}$. 
We introduce the dimensionless quantity $\bar f n_{\uparrow}/\epsilon_{\rm F}$ and, in the following, refer to it as the \textit{mediated interaction coefficient}.
In equations~\eqref{Interaction}-\eqref{eq:QPenergyExperiment} we neglect the direct interaction between the bare impurities, 
since the latter is negligible in our experiment (Methods).

\begin{figure}[t!]
\centering
\includegraphics[width= 88mm]{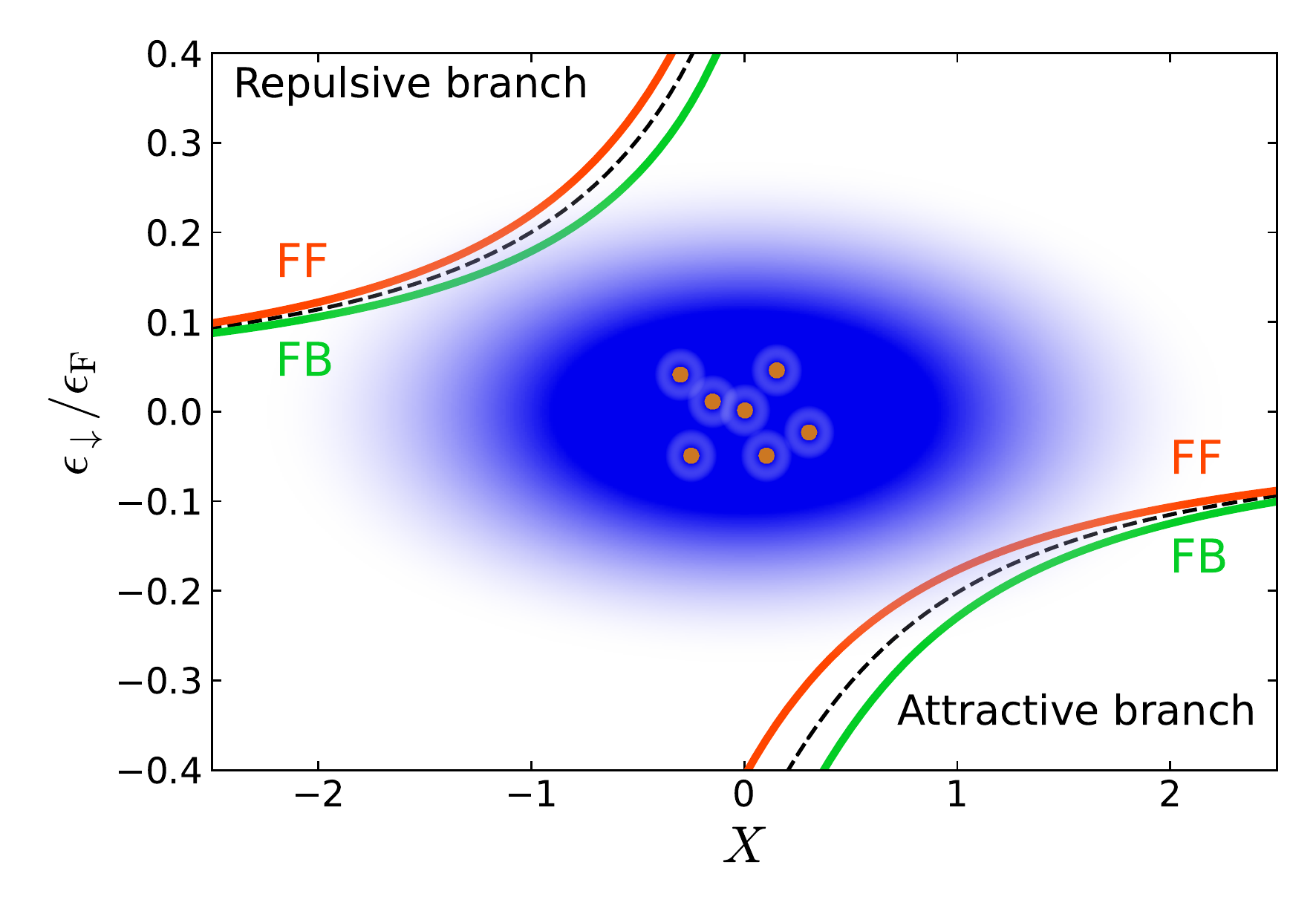}
\caption{\label{fig:fig1}\textbf{Polaron energy including mediated interactions}.
The energy of attractive and repulsive polarons is presented as a function of the dimensionless interaction parameter $X$ according to equation~\eqref{eq:QPenergyExperiment} using the static limit for the polaron-polaron mediated interaction defined in equation~\eqref{Interaction2}. The results are shown for the limit of a single impurity (i.e., $\mathcal{C}=0$, black dashed lines), and for impurity concentration $\mathcal{C} = 0.5$ in a Fermi-Bose (FB, solid green lines) and Fermi-Fermi (FF, solid red lines) Li-K mixture. 
}
\end{figure}

In Fig.~\ref{fig:fig1} we illustrate the interaction dependence of the attractive and repulsive polaron energies according to equation~(\ref{eq:QPenergyExperiment}), with $\bar f=f_0$ given by equation~\eqref{Interaction2},
for three exemplary cases: single impurity (dashed black lines) and  for finite concentration for bosonic (green solid lines) and fermionic (red solid lines) impurities. The figure also highlights that the mediated interaction effect (exaggerated by choosing a large concentration $\mathcal{C}=0.5$) is rather weak, which makes its observation very challenging in an actual experiment. 

The main motivation of our experimental work is to use the unique level of control offered by atomic quantum gases to explore the interaction between quasi-particles systematically. 
This includes testing the behaviour of the sign of the mediated interaction, which is predicted to be independent of the attractive or repulsive character of the impurity-medium interaction, but which should undergo a sign reversal with the impurity quantum statistics.
More generally, we will test the validity of Landau's quasi-particle theory for increasing impurity concentration and variable interaction strength.

Our experiments are carried out with a mixed-species system, which consists of a trapped spin-polarized FS of $^6$Li atoms in which either bosonic or fermionic impurity atoms are immersed ($^{41}$K or $^{40}$K). 
The possibility to change the impurity statistics from bosonic to fermionic, while keeping all the other features nearly identical, is a key feature of our system, allowing for a direct observation of the effect of the impurity statistics in the polaron problem. Moreover, we benefit from the K-Li mass imbalance ($m_\downarrow/m_\uparrow \approx 6.8$) in several ways.
First, it allows to prepare a system with high impurity concentration where the fermionic medium is deeply degenerate while the impurities are kept in the thermal regime (see Supplementary Information). This avoids complications arising for degenerate impurities, such as energy shifts resulting from Pauli blocking for fermionic impurities \cite{Giraud2012ibp, Schirotzek2009oof, Scazza2017rfp}, or phase separation or collapse when a Bose-Einstein condensate of bosonic impurities is formed \cite{Fritsche2021sab}. 
Moreover, the thermal energy distribution of impurities is insensitive to the difference between their bare and their effective mass, the effects of which can, in the degenerate regime, dominate over mediated interactions \cite{Scazza2017rfp} (see Supplementary Information for further details).

As a further benefit, the strong Fermi pressure of the lighter atoms of the medium results in a spatial extent of the FS that is large compared with the cloud size of the heavier impurity atoms. This effect (further enhanced by the tighter optical trapping potential for K in the near-infrared light \cite{Kohstall2012mac}) leads to the favorable situation that the impurities in the FS experience a nearly homogeneous medium with an essentially constant chemical potential.

The basic idea of our probing method, which has been introduced in previous work \cite{Kohstall2012mac, Fritsche2021sab}, is an \textit{injection} scheme based on the radio frequency (RF) transfer from an impurity spin state K$\ket{\text{0}}$ to a state K$\ket{\text{1}}$. The initial state K$\ket{\text{0}}$ is essentially non-interacting with the medium.
The target state K$\ket{\text{1}}$, instead, features tunability of the $s$-wave interaction with the fermionic medium via a magnetically controlled interspecies Feshbach resonance (FR) (Methods). 
In both mixtures (FB and FF), direct interaction between the impurities is negligible since the intraspecies $s$-wave scattering is off-resonant for bosonic $^{41}$K and absent for fermionic $^{40}$K (Methods). 
We note that in real experiments the normalized single polaron energy 
$\epsilon_{\downarrow}^0/\epsilon_{\rm F}$
and the mediated interaction coefficient $\bar f n_\uparrow/\epsilon_{\rm F}$ depend not only on the parameter $X$, but also on the mass ratio and an additional parameter related to the resonance width \cite{Chin2010fri, Massignan2014pdm}. Since in the two mixtures used in our experiment the values of these parameters are almost identical, the differences in mass ratio and FR have a minor effect, so that the impurity quantum statistics is the essential difference.

The starting point of our experiments is a mixture of roughly $10^5$ $^6$Li and $10^4$ $^{41}$K or $^{40}$K atoms. While the Li atoms are always kept in the lowest hyperfine spin state, the K atoms are initially prepared in a weakly interacting ancillary state, K$\ket{\text{anc}}$, from which a fraction of atoms is RF transferred to K$\ket{\text{0}}$ in order to vary the impurity concentration (Methods). The atoms are trapped in a 1064-nm crossed-beam optical dipole trap in the presence of a magnetic field close to the FR. The two species are in thermal equilibrium. While the Li FS is degenerate, the K impurities remain in the thermal regime.

In order to take the residual inhomogeneity of our system into account, we introduce effective quantities defined by averaging on the spatial extent of the K cloud. In particular we introduce the K-averaged atom number densities, $\bar{n}_\text{Li}$ and $\bar{n}_\text{K}$, for the two species, and the effective Fermi energy $\epsilon_\text{F}$ of the Li FS (Methods).
These quantities and the temperature of the sample, $T$, are obtained from separate measurements performed before each main polaron measurement. Typical values of the effective Li density, effective Fermi energy and reduced temperature are similar for the two mixtures. In particular we have
$\bar{n}_\text{Li}\approx 1.5 \times 10^{12}$ cm$^{-3}$, $\epsilon_\text{F}/h \approx 16$~kHz, $T/T_\text{F} \approx 0.15$ for the FB and $\bar{n}_\text{Li}\approx 2 \times 10^{12}$ cm$^{-3}$, $\epsilon_\text{F}/h \approx 20$~kHz, $T/T_\text{F} \approx 0.25$ for the FF mixture.

We implement the RF-spectroscopic probing scheme by applying a 1-ms Blackman-shaped pulse. The pulse duration is chosen as a compromise between the spectral width of the RF pulse ($\sigma_\text{RF} \approx 0.7$ kHz) and the shortest polaron lifetime (of the order of few ms for strong repulsion \cite{Kohstall2012mac, Fritsche2021sab}). The pulse intensity is adjusted to obtain a resonant $\pi$-pulse in the absence of Li atoms. We vary the pulse frequency $\nu$ in order to probe the polaron spectrum. Our spectroscopic observable is the number $N_{\text{K} \lvert 0\rangle }$ of atoms remaining in K$\ket{\text{0}}$ after the RF pulse (Methods). We define the detuning $\Delta E = h(\nu_0 - \nu)$ for the FB case and $\Delta E = h(\nu - \nu_0)$ for the FF case, where $\nu_0$ is the bare frequency of the  K$\ket{\text{0}}$ to  K$\ket{\text{1}}$ transition in the absence of Li. 
The opposite sign in the definitions of $\Delta E$ takes into account that in our FB mixture K$\ket{\text{0}}$ has a higher energy than K$\ket{\text{1}}$ and vice versa in our FF mixture. For both cases, the polaron spectrum is then represented by $N_{\text{K} \lvert 0\rangle }(\Delta E/\epsilon_{\rm F})$.

\begin{figure}[t!]
\includegraphics[width = 1\linewidth]{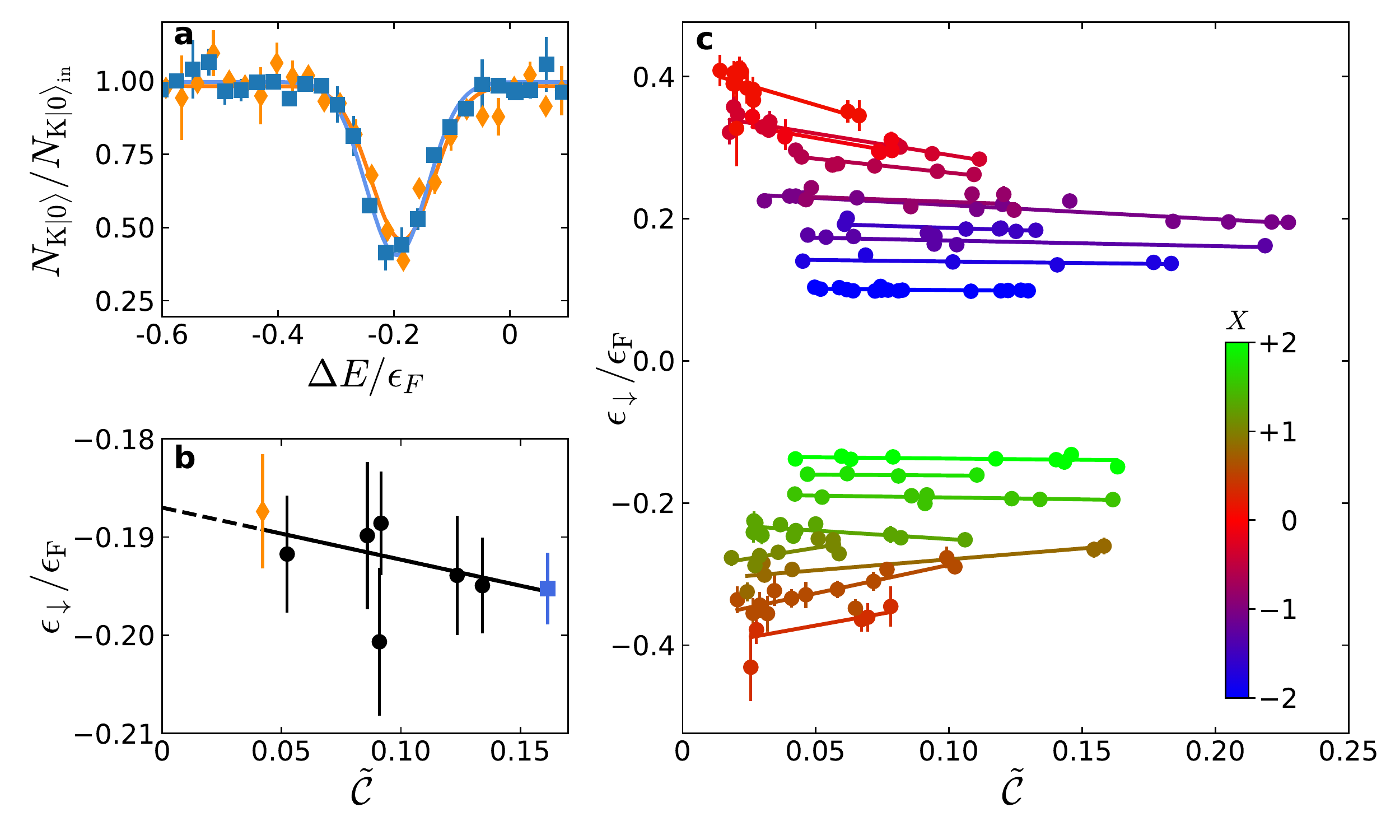} \\
\caption{ \label{fig:fig2} \textbf{Dependence of the polaron energy on the impurity concentration.} 
For the FB case, the main steps of measurements and data analysis are illustrated. \textbf{a}, Two exemplary spectroscopy signals (normalized to the initial atom number) taken at $X = 0.98$ for different values of the interacting impurity concentration (blue squares $\tilde{\mathcal{C}} = 0.16$, orange diamonds $\tilde{\mathcal{C}} = 0.04$). The solid lines are fits with a Gaussian function on a linear background (the latter being negligibly small in the present data). The error bars represent the standard errors from typical 5-6 measurement repetitions. \textbf{b}, Polaron energy as a function of impurity concentration for $X = 0.98$, the blue squares and the orange diamonds correspond to the exemplary spectra presented in panel \textbf{a}. The black line represents a linear fit to the data with the dashed line showing the extrapolation to zero density. \textbf{c}, Polaron energy as a function of impurity concentration for different values of the interaction parameter $X$. From center to top (blue to red) increasing repulsion, from center to bottom (green to red) increasing attraction. Statistical uncertainties are evaluated taking into account fit uncertainties from analyzing the spectra and errors on Fermi energy.}
\end{figure}

In Fig.~\ref{fig:fig2}a we show a typical RF transfer signal, recorded for bosonic $^{41}$K impurities, for two different impurity concentrations. The resonant dip corresponds to the polaron peak in the quasi-particle spectrum and its position reveals a small concentration-dependent energy shift, which we attribute to mediated interactions. We analyze the signal by applying a heuristic fit model, described in Methods, from which we extract the polaron energy peak position, which we identify with $\epsilon_\downarrow$. From the fits we also extract the maximum transferred fraction of impurities from K$\ket{\text{0}}$ to K$\ket{\text{1}}$, $\mathcal{T}_{\text{max}}$. Based on experimental observables, we introduce the  \textit{effective impurity concentration} in the interacting state as $\tilde{\mathcal{C}} = \mathcal{C}_{\text{K} \lvert 0\rangle }  \times \frac{1}{2}\mathcal{T}_{\text{max}}$, where $\mathcal{C}_{\text{K} \lvert 0\rangle } = \bar{n}_{\text{K} \lvert 0\rangle }/\bar{n}_{\text{Li}}$ is the concentration of the noninteracting K$\ket{\text{0}}$ atoms, given by the number density ratio of the two species. The factor $\frac{1}{2}$ in our definition of $\tilde{\mathcal{C}}$ arises from averaging the number of impurities, which are gradually injected from 0 to the final value during the RF pulse \cite{Fritsche2021sab}.

In a second step of our data analysis we consider the dependence of the polaron energy $\epsilon_\downarrow$ on the impurity concentration $\tilde{\mathcal{C}}$ for fixed values of $X$. In Fig.~\ref{fig:fig2}b we present an example of the corresponding procedure for the FB mixture, considering $X\approx 1$. The orange diamond and the blue square represent the polaron energy extracted from the spectra in Fig.~\ref{fig:fig2}a where $\tilde{\mathcal{C}} \approx 0.04$ and $\tilde{\mathcal{C}}\approx 0.16$, respectively. We assume a linear dependence to fit the observed behaviour (black line). From the obtained line we can extract the mediated interaction coefficient 
$\bar f n_\downarrow /\epsilon_{\rm F} $ from the slope, and the normalized single impurity polaron energy $\epsilon^0_\downarrow /\epsilon_{\rm F}$ from extrapolation to zero. In Fig.~\ref{fig:fig2}c we present the measured polaron energies and the corresponding fits of the concentration dependence for different $X$ values for the FB case. 

For fermionic $^{40}$K impurities (FF mixture) we proceed in essentially the same way, apart from details in the preparation process and a generally more noisy signal (Methods). The experimental conditions for the RF-spectroscopic measurements stay close to the case of bosonic $^{41}$K impurities (FB mixture), and the data analysis protocol is identical. The Extended Data Fig.~\ref{fig:extended_fig1} is the analogue of Fig.~\ref{fig:fig2} for the FF case. 
For all our recorded data, the concentration-dependence of the polaron energy is consistent with a linear behaviour, as expected from equation~\eqref{eq:QPenergyExperiment}, at every interaction strength for both the FB and the FF case (see Fig.~\ref{fig:fig2} and Extended Data Fig.~\ref{fig:extended_fig1}).

\begin{figure}[t!]
\centering
\includegraphics[width= 88mm]{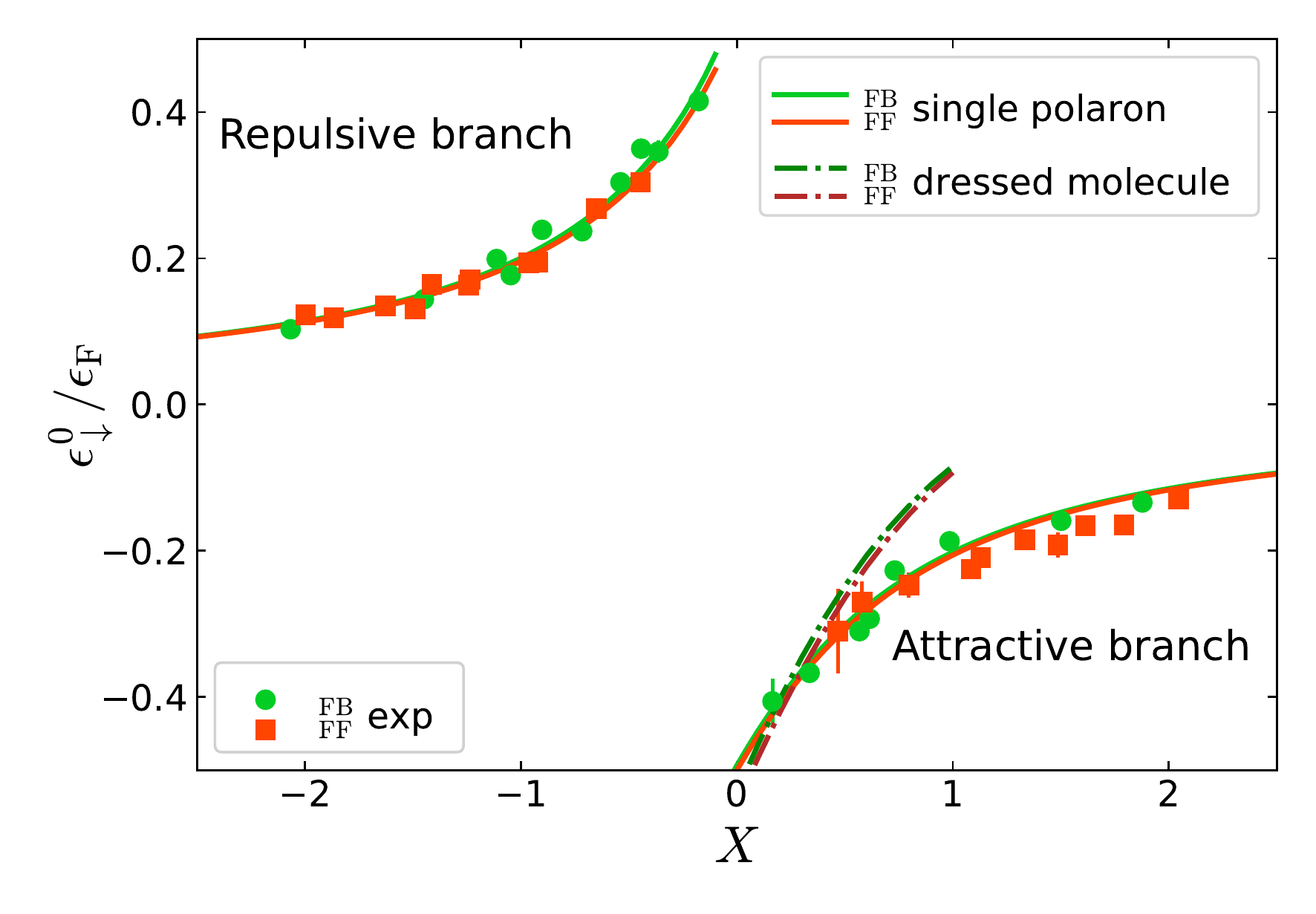} 
\caption{\label{fig:fig3} \textbf{Polaron energy in the single-impurity limit.}
The experimental results for $\epsilon^0_\downarrow/\epsilon_{\rm F}$ 
for the FB (green circles) and the FF (red squares) mixture are compared with the theoretically expected values for a single polaron (green and red solid lines for FB and FF, respectively). In addition we show the energy of the dressed molecules (green and red dash-dotted lines for FB and FF, respectively). Note that the theory lines are almost identical for the FB and the FF case, so that they overlap to a large extent. Error bars for $\epsilon^0_\downarrow/\epsilon_{\rm F}$ correspond to the uncertainties of the linear fit and the errors on the Fermi energy. Error bars for $X$, smaller than the symbol size, represent the standard errors for each set of measurements. 
}
\end{figure}

We can now compare our experimental results with the theoretical prediction of equation~\eqref{eq:QPenergyExperiment}. As an important benchmark, we first confront the zero-concentration result $\epsilon_\downarrow^0$, as obtained from our linear fits, with 
the first term in equation~\eqref{eq:QPenergyExperiment}. This comparison is reported in Fig.~\ref{fig:fig3}, demonstrating excellent 
agreement between experimental results and polaron theory \cite{Chevy2006upd, Massignan2011rpa} in the whole range of interaction strength explored in our experiment. Minor deviations observed for fermionic impurities for $X\gtrsim1$ can be explained considering the finite temperature of our samples (see details
in the Supplementary Information).
The overall agreement with a well-established limiting case represents a validity check for our measurements and, in particular, for the assumed linear dependence of $\epsilon_\downarrow$ on\ $\tilde{\mathcal{C}}$. The impurity quantum statistics cannot play a role in the single-impurity limit, which is in accordance with our experimental observations.

\begin{figure}[t!]
\centering
\includegraphics[width= 88mm]{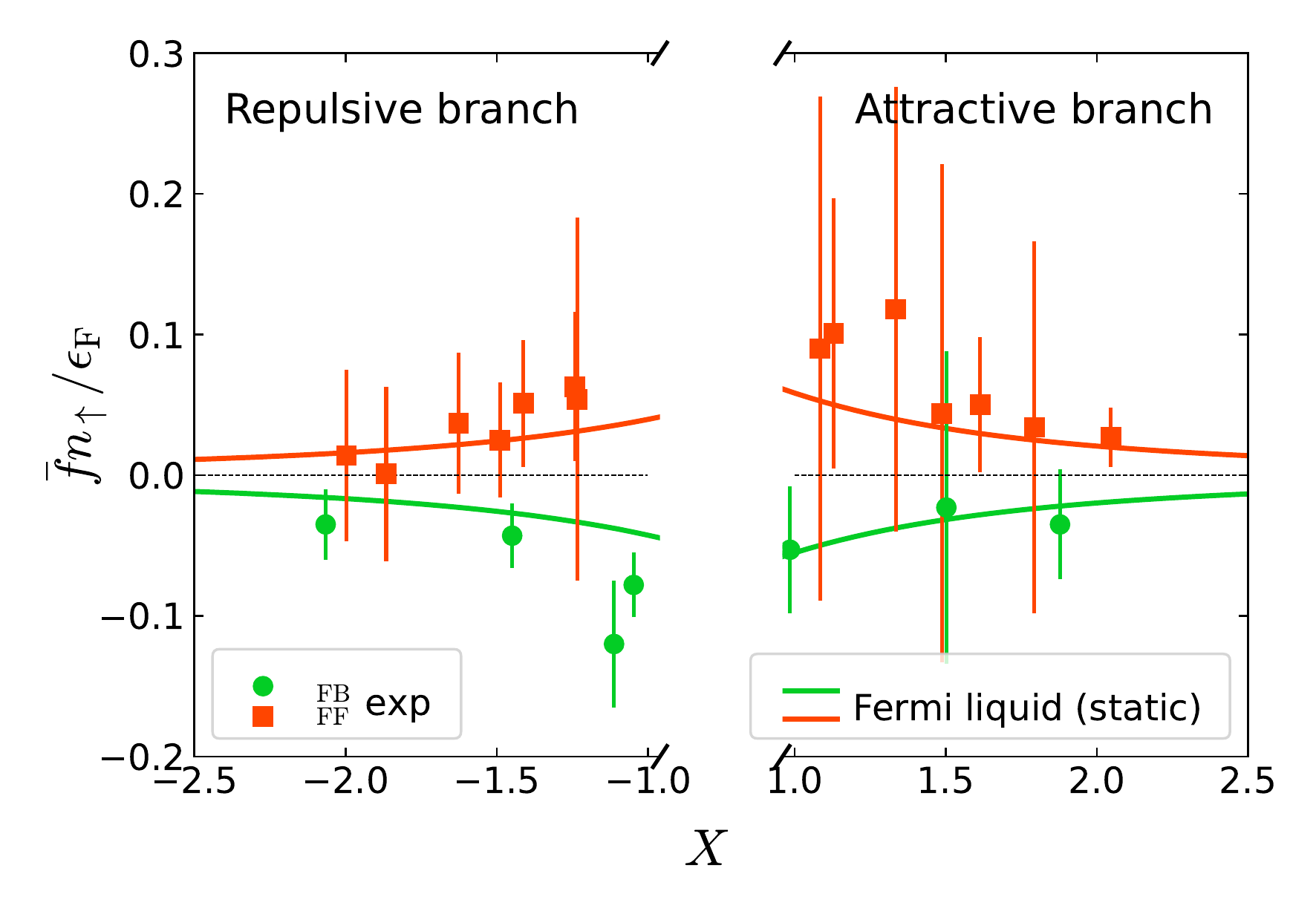} 
\caption{\label{fig:fig4} \textbf{Mediated interaction coefficient in the regime of moderate impurity-medium interactions.}
The experimental results for $\bar f n_\uparrow /\epsilon_{\rm F}$
for FB (green circles) and FF (red squares) are compared with equation~\eqref{eq:QPenergyExperiment} with $\bar f$ given by the static limit of the Fermi liquid theory of equation~\eqref{Interaction2} (green and red solid lines refer to FB and FF cases, respectively) for interaction values $\lvert X \rvert \gtrsim1$. Error bars on $\bar f n_\uparrow /\epsilon_{\rm F}$ correspond to the uncertainties of the linear fit and the errors on the Fermi energy. Error bars on $X$, smaller than the symbol size, represent the standard errors for each set of measurements. }
\end{figure}

\begin{figure}[t!]
\centering
\includegraphics[width= 88mm]{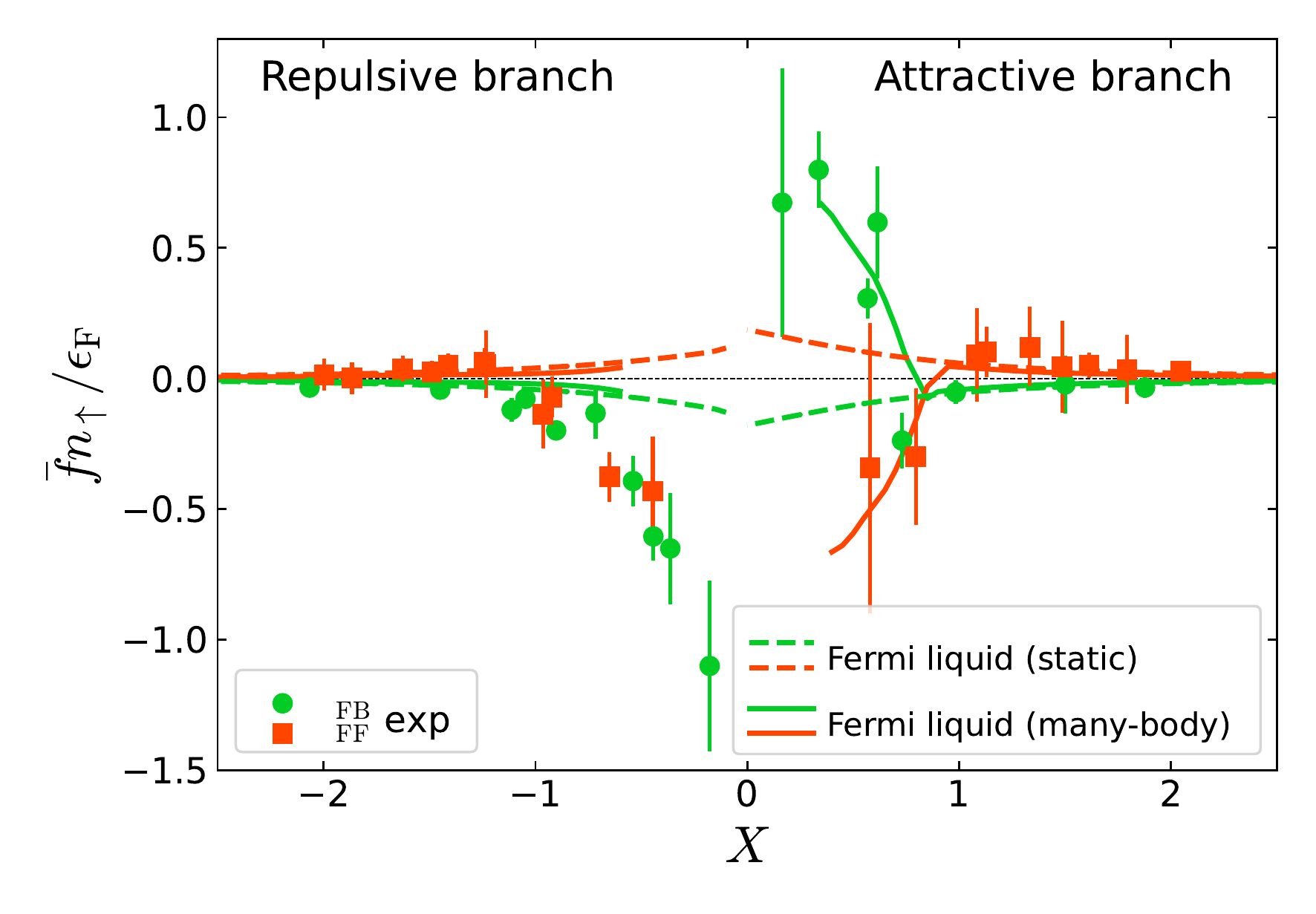} 
\caption{\label{fig:fig5} \textbf{Mediated interaction coefficient across the resonance.}
The experimental results for $\bar f n_\uparrow /\epsilon_{\rm F}$ for the FB (green circles) and the FF (red squares) mixture are compared with the static Fermi liquid theory of equation \eqref{Interaction2} (dashed green and red lines for FB and FF, respectively) and with the corresponding microscopic many-body result (see details in the Supplementary Information), assuming a density of molecules equal to $2.5$ times that of the impurities (solid green and red lines for FB and FF, respectively). Error bars on $\bar f n_\uparrow /\epsilon_{\rm F}$ correspond to the uncertainties of the linear fit and the errors on the Fermi energy. Error bars on $X$, smaller than the symbol size, represent the standard errors for each set of measurements. }
\end{figure}

Let us turn our attention to the main quantity of interest. In our linear fits, the slope yields the mediated interaction coefficient $\bar f n_\uparrow /\epsilon_{\rm F}$.
Most importantly, the experimental results displayed in Fig.~\ref{fig:fig4} for the regime of weak-to-moderate interactions ($\lvert X \rvert \gtrsim 1$) clearly show the expected behaviour. 
For bosonic impurities, the downshift with increasing concentration reveals attraction ($\bar f<0$) on both sides of the resonance. In contrast, for fermionic impurities, the observed interaction between the polarons is repulsive ($\bar f>0$) on both sides of the resonance. Our observations thus highlight the essential difference between bosonic and fermionic impurities, which is the opposite sign of the mediated polaron-polaron interaction shift.
Moreover, the observed strength of the mediated polaron-polaron interaction agrees, in the regime of weak and moderate interactions, with the static Fermi liquid result for $\bar f$ given by equation~\eqref{Interaction2}.

In Fig.~\ref{fig:fig5} we report the observed dependence of the mediated interaction coefficient across the full resonance, exploring in particular the strongly interacting regime $-1 \lesssim X \lesssim 1$.
For strong interactions, we observe striking deviations from the prediction of equation~\eqref{Interaction2}, both regarding the magnitude and sign of $\bar f$. 
On the attractive side of the resonance ($X>0$), a remarkable sign change of $\bar f$ is observed for both bosonic and fermionic impurities for $X \lesssim 1$, so that the interaction becomes repulsive or attractive, respectively. 
When the resonance is approached from the repulsive side ($X<0$), the energy shift becomes negative corresponding to large and negative  values of $\bar f$. 
This appears independently of the impurity quantum statistics, in contrast to the observations for moderate interaction strength and for the attractive side.
  
The solid lines in Fig.~\ref{fig:fig5} show the  prediction of a microscopic 
many-body calculation based on the ladder approximation generalised to non-zero impurity concentrations (Methods). When expanded to linear order in the impurity density, this naturally yields equation~\eqref{eq:LandauQPenergy} for 
the polaron energy together with a microscopic expression for 
the interaction $f_{\mathbf k,\mathbf k'}$~\cite{Bastarrachea2021aar,Bastarrachea2021pia}. This theory recovers equation~\eqref{Interaction2} for weak impurity-medium interaction and low momenta, and generalises this perturbative result by including strong two-body correlations and the momentum dependence of the interaction, the thermal distribution of polarons, retardation, as well as the presence of Feshbach molecules. 
We see that it recovers the experimental results on the attractive 
side ($X>0$) including the observed sign change in the strong coupling regime. The theory attributes this sign change to a significant thermal population of Feshbach molecules as their energy approaches that of the polaron close to resonance~\cite{Punk2009ptm,Trefzger2012iia, Massignan2012pad}. This can be understood from the fact that the coupling to higher energy molecules gives rise to a negative energy shift of the polarons~\cite{Bastarrachea2021aar}. For fermionic 
impurities, these Feshbach molecules are bosons and it follows that a thermal population will lead to Bose stimulation increasing this negative energy shift. This corresponds to an \emph{attractive} polaron-molecule 
interaction that counteracts the repulsive polaron-polaron interaction for fermionic impurities, giving the observed decrease in the polaron energy for the FF mixture at strong coupling. An analogous argument yields a \emph{repulsive} polaron-molecule interaction for bosonic impurities, resulting in the 
abrupt increase in the calculated polaron energy in the FB mixture 
shown in Fig.~\ref{fig:fig5}. We have assumed a molecule density $2.5$ times the impurity density to calculate the curves in Fig.~\ref{fig:fig5}, 
which is consistent with the experimental conditions, and we have  restricted the plotted range to the regions where the polaron residue exceeds 0.5. 

The pronounced  decrease in the energy of repulsive polarons in the strongly interacting regime $-1\lesssim X<0$, which is independent of the impurity 
statistics, cannot be explained by our theory. 
We speculate that this may be due to a breakdown of the quasi-particle 
picture for strong coupling.  Indeed, the quasi-particle residue is small for strong coupling and the repulsive polaron becomes strongly damped~\cite{Massignan2014pdm}, raising  questions on a description in terms of quasi-particles. Our results therefore motivate further 
studies exploring  the intricate interaction physics in the strongly interacting regime, which may involve intriguing physics well beyond the Fermi liquid 
paradigm. Moreover, it would be interesting to explore further the breakdown of the quasi-particle picture with increasing 
concentration. Indeed, the underlying physics changes for  
$\mathcal{C} \approx 1$ to nearly balanced FB \cite{Fratini2010pac, Ludwig2011qft, Yu2011sco, Duda2023tfa} and FF \cite{Gubbels2013ifg, Pini2021bmf} mixtures, 
and, for $\mathcal{C} \gg 1$, the roles of the impurities and the medium are reversed \cite{Fritsche2021sab}. \\

We acknowledge stimulating discussions with J.~Dalibard, M.~Parish, J.~Levinsen, and Z.~Wu.

The project has received partial funding from the European Research Council (ERC) under the European Union's Horizon 2020 research and innovation program (grant agreement No. 101020438 - SuperCoolMix)

P. M. acknowledges support by grant PID2020-113565GB-C21 from MCIN/AEI/10.13039/501100011033, by grant 2021 SGR 01411 from the Generalitat de Catalunya, and by the {\it ICREA Academia} program. 

G.~M.~B.\ acknowledges support by the Danish National Research Foundation through the Center of Excellence “CCQ” (Grant agreement no.: DNRF156).

The authors declare no competing interests.




\newpage

\appendix

\section*{Methods}

\subsection*{Spin states}

For the RF spectroscopic measurements, the $^6$Li atoms forming the spin-polarized FS are always kept in the lowest hyperfine spin state $F, m_F = (1/2, +1/2$). In the preceding preparation process, two-component spin mixtures are exploited for evaporative cooling.

As the starting point for the measurements, the K atoms are prepared in an `ancillary' state K$\ket{\rm anc}$, which for $^{41}$K corresponds to the third-lowest spin state (1, +1) and for $^{40}$K to the lowest state (9/2, $-9/2$). In our spectroscopic injection scheme, the initial state K$\ket{\rm 0}$ is represented by the state (1, 0) for $^{41}$K and by the state (9/2, $-7/2$) for $^{40}$K. The final Feshbach-resonant state K$\ket{\rm 1}$ corresponds to ($1, +1$) for $^{41}$K and to the state (9/2, $-5/2$) for $^{40}$K.

\subsection*{Sample preparation}

The Li and K atoms are first loaded from a Zeeman-slowed atomic beam into a dual-species magneto-optical trap and then transferred into a single-beam optical dipole trap (ODT1), which is derived from a 200-W fiber laser at a wavelength of 1070\,nm. During the transfer process we apply grey-molasses cooling on the D1 line of Li, which reduces the temperature to about 50$\mu$K and polarizes the majority of Li atoms into the lowest hyperfine spin state ($1/2, +1/2$) \cite{Burchianti2014eao, Fritsche2015MASTER}. Mixing the lowest two spin states with a resonant RF, we then produce a 50/50 mixture in the lowest two states ($1/2, \pm1/2$). We hold the sample for a time of typically 0.5\,s (3\,s) in the FB (FF) case, in which spin relaxation induced by interspecies collisions polarizes the K atoms into the state K$\ket{\rm anc}$ \cite{Spiegelhalder2010aop, Lous2017toa}. This spin relaxation process is optimized at a magnetic field of $\sim$200\,G for $^{41}$K and $\sim$15\,G for $^{40}$K. 

Evaporative cooling is implemented by ramping the power of the ODT1 down to zero while ramping down the power of a crossed optical dipole trap (ODT2), which is turned on at maximum power together with the ODT1 at the loading stage. The ramp is performed in $\sim$ 5\,s and a wait time of 1 s is added to reach thermal equilibrium between the two species. During the evaporation stage the K atom are sympathetically cooled by the Li atoms, which are in a mixture of ($1/2, 1/2$) and ($3/2, -3/2$) states at 485 G for the FB case and in a mixutre of ($1/2, 1/2$) and ($1/2, -1/2$) states at 923 G for the FF case.
The final power of the ODT2 is adjustable in order to reach the desired temperature of the atomic clouds and we rise it in the final stage of the preparation process to reach the same trapping frequencies for each experimental run, minimizing the effect of optical shifts of the FR induced by the trap light (see Sup. Mat. of \cite{Cetina2015doi}). The geometrically averaged trap frequencies are 117\,Hz for $^{41}$K and 193\,Hz for Li for the FB mixture, and  158\,Hz for $^{40}$K and 271\,Hz for Li for the FF mixture. The ODT2 features an aspect ratio of about seven, with the weak axis oriented horizontally. The differential gravitational sag of both species amounts to about 3\,$\mu$m and can be neglected since the FS is much larger.

After evaporative cooling, in order to obtain a fully polarized sample of Li in the ($1/2, +1/2$) state, we remove the atoms in the second spin state of Li by applying a resonant 10-$\mu$s light pulse at 567\,G (1180\,G) in the FB (FF) case.

Then the magnetic field is ramped close to the FR of interest near 335\,G for the FB mixture and 155\,G for the FF mixture. At this point we switch to a different set of coils, which facilitates precise magnetic field control to the level below a few mG. The trap contains a thermalized sample of roughly 10$^5$ Li atoms in the lowest hyperfine spin state and $10^4$ K$\ket{\rm anc}$ atoms. The temperature $T$ is 102 -- 137\,nK in the FB case and 190 -- 285\,nK in the FF case. At these temperatures, the FS is deeply degenerate with $T/T_F \approx 0.15$ and $T/T_F \approx 0.25$ for the FB and FF mixture, respectively. In contrast, the K impurities remain in the thermal regime.

In order to vary the impurity concentration in the initial spectroscopy state, we transfer the desired amount of atoms from the ancillary state K$\ket{\text{anc}}$ to K$\ket{\text{0}}$ using a RF pulse. In this way, keeping the total number of K atoms constant in the cooling process, we ensure identical starting conditions for measurements with variable impurity concentration.

\subsection*{Detection}
\label{sec:detection}

After the RF pulse that is used for probing the polaron spectrum we switch off the ODT2 and let the atomic cloud expand for a typical time of 1~ms and we detect the atoms by using state-selective absorption imaging. In the case of the FB mixture we can image K$\ket{\text{0}}$ and K$\ket{\text{1}}$ directly, using nearly closed optical transitions. In the FF case, however, because of the larger hyperfine structure of $^{40}$K in combination with the lower magnetic field, optical transitions for the detection of K$\ket{\text{0}}$ and K$\ket{\text{1}}$ are leaky, which compromises the signal-to-noise ratio. Here, we apply a resonant light pulse in order to remove K$\ket{\text{1}}$ in 10 $\mu$s and apply a 96-$\mu$s RF $\pi$-pulse transferring atoms from K$\ket{\text{0}}$ to K$\ket{\text{anc}}$, which permits imaging via a closed optical transition. In this way, we derive our spectroscopic signal from atoms remaining in state K$\ket{\text{0}}$. For the FB case, to facilitate a direct comparison, we also define the corresponding spectroscopic signal based on atoms remaining in K$\ket{\text{0}}$.

We determine the temperature of the mixture by ballistic expansion of the thermal K cloud after releasing it from the trap, both for the FB and FF cases. The Fermi energy of the Li cloud and the concentration of the noninteracting impurities are obtained from separate measurements in which we record the distributions of the Li and K atoms with the same initial conditions as in the polaron spectrum acquisition. These measurements are performed before each polaron spectrum is taken. \\

\subsection*{Effective quantities}

In order to take spatial inhomogeneities in our harmonically trapped mixture into account, we introduce effective quantities defined by averaging over the spatial extent of the K cloud, as in previous works \cite{Cetina2016umb, Fritsche2021sab}. We introduce the K-averaged atom number densities, $\bar{n}_\text{Li}$ and $\bar{n}_\text{K}$, for both species, namely
\begin{align}
  \bar{n}_\text{Li,K} = \frac{1}{N_\text{K}}\int n_\text{Li,K}(\textbf{r}) n_\text{K}(\textbf{r}) d^3\textbf{r},
\end{align}
with $n_\text{Li,K}(\textbf{r})$ being the local number densities of Li and K at a position $\textbf{r}$, respectively. Similarly we define the effective Fermi energy as 
\begin{align}
  \epsilon_\text{F} = \frac{1}{N_\text{K}}\int E_\text{F}(\textbf{r}) n_\text{K}(\textbf{r})d^3\textbf{r},
\end{align}
where the local Fermi energy at position $\textbf{r}$ is given by 
\begin{align}
  E_\text{F}(\textbf{r}) = \frac{\hbar^2(6\pi^2n_\text{Li}(\textbf{r}))^{2/3}}{2m_\text{Li}}.
\end{align}
From this we define the effective Fermi wave number as ${k_\text{F} = \sqrt{2m_\text{Li}\epsilon_\text{F}}/\hbar}$ \cite{Kohstall2012mac}. 
We emphasize that, owing to the much smaller size of the K cloud, the effect of inhomogeneity in $n_\text{Li}$, and thus in $E_\text{F}$, is small and, for $n_\text{Li}$, it amounts to about $10\%$.

\subsection*{Control of interactions}

The $s$-wave interaction of bosonic or fermionic impurity atoms in the spectroscopy state K\ket{\rm 1} with the $^6$Li atoms forming the FS can be controlled by means of interspecies Feshbach resonances. The $s$-wave scattering length follows the standard expression
\begin{equation}
a(B) = a_{\rm {bg}} \left( 1 - \frac{\Delta}{B-B_0}\right) ,
\end{equation}
where $B_0$ represents the resonance center, $\Delta$ denotes the width, and $a_\text{bg}$ is the background scattering length.

For the $^{41}$K-$^6$Li resonance (FB case), the relevant parameter values are \cite{Fritsche2021sab, Lous2018pti}: $B_0 = 335.080(1)$\,G, $\Delta = 0.9487$\,G, and $a_\text{bg} = 60.865\,a_0$. The closed-channel dominated resonance \cite{Chin2010fri} is further characterized by a differential magnetic moment $\delta \mu/h = 2.660(8)$\,MHz/G. For the range parameter, as introduced in Ref.~\cite{Petrov2004tbp}, this corresponds to the value $R^* = 2241(7)\,a_0$ or, if expressed in relation to the Fermi wave number in a dimensionless way, to $\kappa_F R^* = 0.54$. 

For the $^{40}$K-$^6$Li resonance (FF case), the relevant parameter values are \cite{Cetina2016umb, Naik2011fri}: $B_0 = 154.7126(5)$\,G, $\Delta = 0.88$\,G, and $a_\text{bg} = 63.0\,a_0$. The resonance is further characterized by a differential magnetic moment $\delta \mu/h = 2.3$\,MHz/G. For the range parameter, this corresponds to $R^* = 2405(63)\,a_0$ or $\kappa_F R^* = 0.62$. Here the reported values of $B_0$ and $R^*$ are determined during the preparation of the experiment reported in this work, following the molecule dissociation technique described in \cite{Fritsche2021sab}. 

We emphasize the very similar character of the resonances in the FB and the FF case. Although they are both closed-channel dominated \cite{Chin2010fri}, their width is large enough to stay in a near-universal interaction regime ($\kappa_F R^* < 1$) with the FS.

For both K isotopes in the initial spectroscopy state K\ket{0} and the ancillary state K\ket{\rm anc}, the interaction with the FS remains weak with scattering lengths of about $60\,a_0$ \cite{Naik2011fri}. This is just enough for thermalization with the medium in the sympathetic cooling process, but for the spectroscopic scheme this background interaction is fully negligible.

We further note that the weak intraspecies interactions between all the K considered spin states (K\ket{0}, K\ket{\rm anc} and K\ket{1}) are characterized by small scattering lengths, of the order of $60\, a_0$ and $170\, a_0$ for the FB \cite{Lysebo2010fra} and FF \cite{Ludewig2012fri} mixtures, respectively. In addition the intraparticle scattering length for the fermionic K$\ket{1}$ is negligible due to Pauli blocking.

\subsection*{Spectroscopic signal and fit}
\label{sec:fit}

Our choice of the number $N_{\text{K} \lvert 0\rangle }$ of atoms remaining in K$\ket{\text{0}}$ after the RF pulse as the spectroscopic observable has the advantage of working equally well for the FB and FF cases. In previous work, we had chosen $N_{\text{K} \lvert 1\rangle }/N_{\text{K}_\text{tot}}$ (number of atoms in K$\ket{\text{1}}$ after the pulse divided by the total atom number) \cite{Kohstall2012mac, Fritsche2021sab} or $(N_{\text{K} \lvert 1\rangle } - N_{\text{K} \lvert 0\rangle })/N_{\text{K}_\text{tot}}$ \cite{Cetina2016umb} to extract the signal. However, the absence of a sufficiently closed optical transition impedes an efficient detection of atoms in K$\ket{\text{1}}$ in particular for the FF mixture. Moreover, possible losses in the transfer process (though not observed) may complicate the interpretation of a signal based on $N_{\text{K} \lvert 1\rangle }$.

We take measurements in a range of about $\pm 35$\,mG around the FR centre. This implies that we need knowledge of our magnetic field on the level of a mG. The main limitation in our system is given by slow drifts in the magnetic field during the measurements. In order to deal with this drift we perform magnetic field calibration measurements before and after each spectrum is recorded and we reject spectra for which the difference of the results of these two measurement is larger than 2\,mG (1\,mG) for FB (FF). Moreover we perform the measurements in the shortest time possible and, as a consequence, we sample the quasi-particle spectrum in a region centered around the polaron peak (typically less than $\epsilon_\text{F}$ wide), not recording the full background.

As a well-established fact, the spectroscopy signal has two component: one due to the polaron and one due to the presence of an underlying incoherent background \cite{Massignan2014pdm, Cetina2016umb,Fritsche2021sab}. In previous work \cite{Kohstall2012mac, Fritsche2021sab}, the broad background was approximated by a Gaussian, so that the complete polaron spectrum could be modeled by a double Gaussian function.
Given the incompleteness of the recorded background in the present measurements, applying a double Gaussian fit to our spectrum is not possible. To circumvent this we fit with a Gaussian combined with a linear background, the latter mimicking the presence of the not fully recorded incoherent background, especially present in the strongly interacting regime and negligible for weak interactions. We checked that fitting selected spectra taken over a wider range with a Gaussian and a linear fit and with a double Gaussian leads to values of the polaron energy that are in each other error bars (see Extended Data Fig.~\ref{fig:extended_fig2}).

\subsection*{Theory}

In order to describe strong coupling effects including the momentum dependence of the effective interaction, a thermal distribution of polarons, and the 
 presence of Feshbach molecules, 
we apply microscopic many-body theory. In this approach, the polaron energy is given by solving 
\begin{equation}
\epsilon_{\mathbf k\downarrow}=\hbar^2 k^2/2m_\downarrow+\Sigma(k,\epsilon_{\mathbf k\downarrow}),
\label{QPeqn}
\end{equation}
 where $\Sigma(k,\omega)$ is the impurity self-energy. To connect this to the Landau form given by equation~\eqref{eq:LandauQPenergy}, we   
 expand the self-energy to linear order in the impurity population writing
$\Sigma(\mathbf k,\omega)\simeq\Sigma_0(\mathbf k,\omega)+\sum_{\mathbf k'}n_{\mathbf k'\downarrow}\partial_{n_{\mathbf k'\downarrow}}\Sigma(\mathbf k,\omega)\vert_{n_{\mathbf k'\downarrow}=0}$
where $\Sigma_0(\mathbf k,\omega)$ denotes its value for a single impurity.
Solving equation~\eqref{QPeqn} to linear order in the impurity population then yields equation~\eqref{eq:LandauQPenergy} with the interaction
\begin{equation}
f_{\mathbf k,\mathbf k'}=Z_{\mathbf k}\left.\frac{\partial \Sigma(\mathbf k,
{\epsilon_{\mathbf k\downarrow}}
)}{\partial n_{\mathbf k'\downarrow}}
 \right\vert_{n_{\mathbf k'\downarrow}=0}.
 \label{MBtoLandau}
 \end{equation}
where $Z_{\mathbf k}=[1-\partial_\omega\Sigma^0(\mathbf k,\omega)_
{\omega=\epsilon_{\mathbf k\downarrow}}
]^{-1}$ is the quasi-particle residue. 
Equation \eqref{MBtoLandau} shows in general how   the effective interaction between quasi-particles can be calculated from microscopic 
many-body theory. 

To proceed, we calculate the self-energy using the ladder approximation, which is remarkably accurate for describing isolated Fermi polarons even for strong interaction~\cite{Massignan2014pdm}.
To compute the polaron-polaron coupling at vanishing temperature and momentum $f_0$ introduced in equation~\eqref{Interaction2} one only needs the single-impurity self-energy, since $\Delta N = -Z\, \partial_{\epsilon_{\text{F}}} \Sigma^0(0,\epsilon_{\mathbf k\downarrow})$.

The evaluation of the general coupling $ f_{\mathbf k,\mathbf k'}$ obtained in equation~\eqref{MBtoLandau} requires instead that one computes the self-energy for non-zero impurity concentration (see details in the Supplementary Information).
This yields an extra term in the self-energy coming from the interaction of the polaron with Feshbach molecules, which becomes significant close to resonance where the molecule energy approaches that of the polaron. 
The theory predicts the molecule-polaron interaction to be attractive/repulsive for fermionic/bosonic impurities~\cite{Bastarrachea2021aar}.

\newpage
\section*{Extended Data Figures}

\begin{figure}[h!]
\vspace{-5mm}
\includegraphics[width = 1\linewidth]{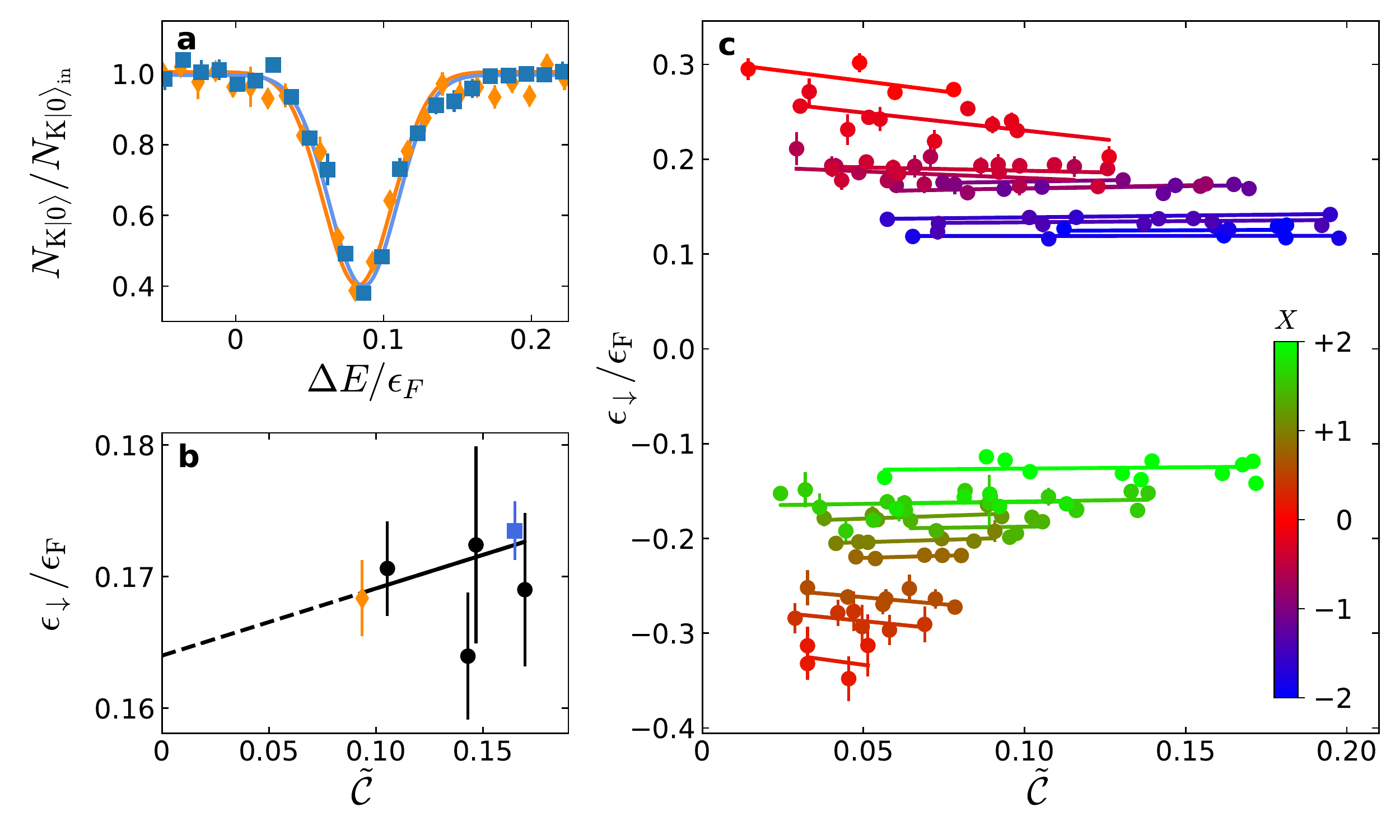}
\caption{ \label{fig:extended_fig1} \textbf{Dependence of the polaron energy on the impurity concentration.} 
For the FF case, the main steps of measurements and data analysis are illustrated \textbf{a}, Two exemplary spectroscopy signals (normalized to the initial atom number) taken at $X = -1.41$ for different values of the interacting impurity concentration (blue squares $\tilde{\mathcal{C}} = 0.17$, orange diamonds $\tilde{\mathcal{C}} = 0.09$). The solid lines are fits with a Gaussian function on a linear background (the latter being negligibly small in the present data). The error bars represent the standard errors from typical 5-6 measurement repetitions. \textbf{b}, Polaron energy as a function of impurity concentration for $X = -1.41$, the blue squared and the orange diamond correspond to the exemplary spectra presented in panel \textbf{a}. The black line represents a linear fit to the data with the dashed line showing the extrapolation to zero density. \textbf{c}, Polaron energy as a function of impurity concentration for different values of the interaction parameter $X$. From center to top (blue to red) increasing repulsion, from center to bottom (green to red) increasing attraction. Statistical uncertainties are evaluated taking into account fit uncertainties from analyzing the spectra and errors on the Fermi energy.
}

\end{figure}

\begin{figure}[t!]
\centering
\includegraphics[width = 1\linewidth]{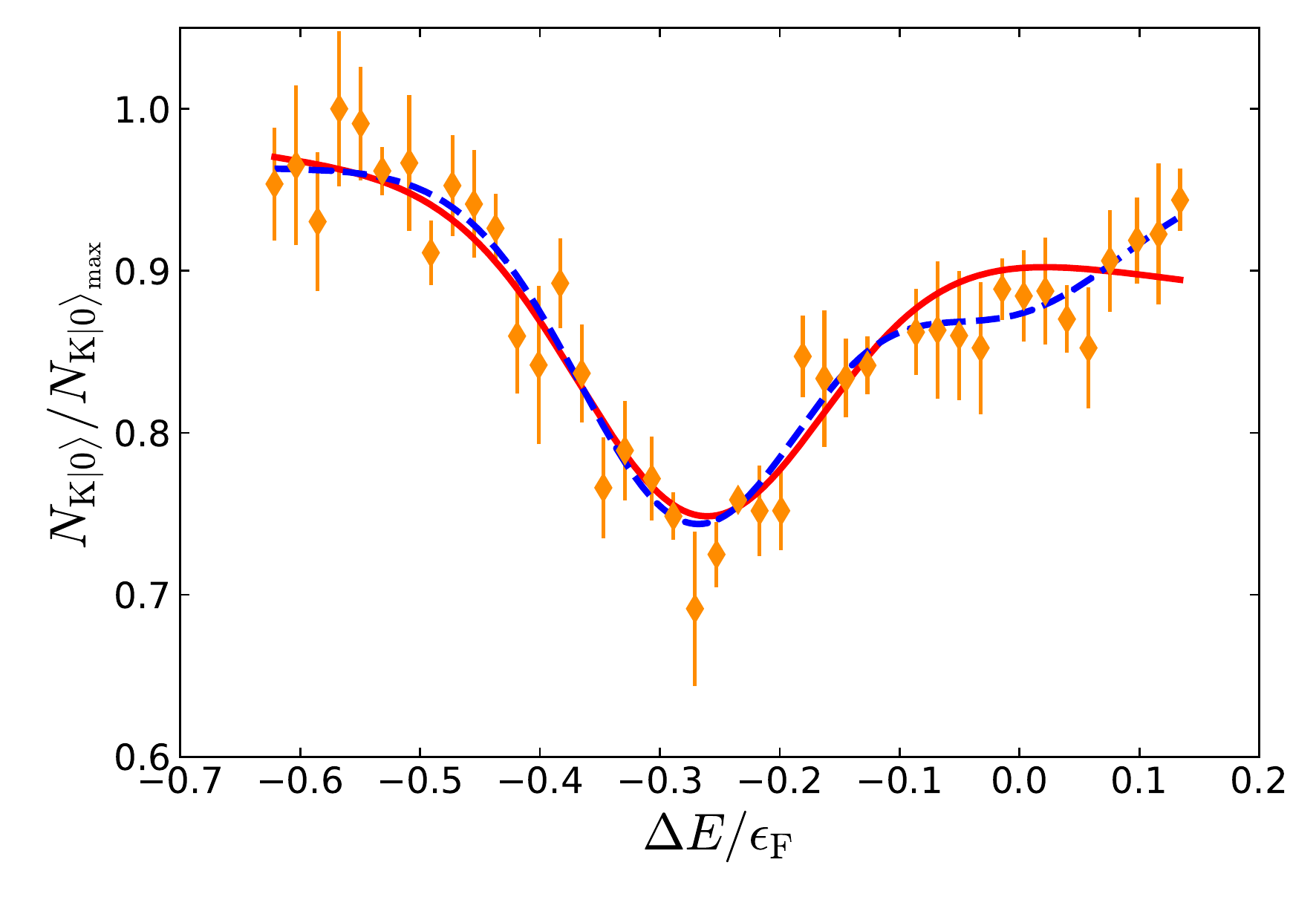}
\caption{ \label{fig:extended_fig2} \textbf{Comparison between two fitting functions}. Spectroscopy signal, normalized to the maximum atom number $N_{\rm K\ket{0}_{\rm max}}$, for an exemplary measurement in the FF case at at $X = 0.75$ and for $\tilde{\mathcal{C}} = 0.08$ fitted by a Gaussian plus linear background (red solid line) and by a double Gaussian (blue dashed line). The resulting dip position are at $\Delta E_{\rm p} = -0.266(7)$ and $\Delta E_{\rm p} = -0.27(1)$, respectively. The error bars represent the standard errors from typical 5-6 measurement repetitions.}
\end{figure}

\end{document}



\title{Supplementary Information for \\ Mediated interactions between Fermi polarons and the role of impurity quantum statistics}

\author{Cosetta Baroni, Bo Huang (\begin{CJK*}{UTF8}{gbsn}黄博\end{CJK*}), Isabella Fritsche, Erich Dobler, Gregor Anich, Emil Kirilov, Rudolf Grimm, Miguel A. Bastarrachea-Magnani, Pietro Massignan, and Georg Bruun}

\maketitle

\section{Impurity-impurity interactions and thermal effects}

In equation (4) of the main text we neglect the effect of the mean energy of the impurities described in Ref.~\cite{Scazza2017rfp}. 
This effect arises from the fact that the initial non interacting impurity, with bare atomic mass $m_{\rm K}$, and the final quasi-particle, with effective mass $m_{\rm K}^*$, show different dispersion relations. Since the radio-frequency pulse bringing the impurities from the non interacting to the interacting state does not change their momentum, the energy of the polaron will be shifted by an amount proportional to the impurity mean energy $\bar \varepsilon$.
The fact that in our system the K impurities are thermal, allows us to neglect this effect.
In order to justify this claim, we carry out the same calculation reported in equation~(S.4) of the Supplemental Material of Ref.~\cite{Scazza2017rfp}: 
\beq
\label{eq_scazza}
\bar\varepsilon(\mathcal{C}) \equiv \bar \varepsilon_{\text{K} \ket{0}}(\mathcal{C}) = \frac{4 \pi}{(2\pi \hbar)^3 N_{\text{K} \ket{0}}(\mathcal{C})} \int_{\mathcal{V}} d{\bf{r}} \int_{0}^{\infty} dp\, p^2 \frac{\varepsilon_{\bf{p},\bf{r}}}{e^{(\varepsilon_{\bf{p},\bf{r}}-\mu(\mathcal{C}))}+1}
\eeq
where $\varepsilon_{\bf{p},\bf{r}} = p^2/2m_K + U({\bf{r}})$ is the single-particle energy of the impurities, $\mu(\mathcal{C})$ their chemical potential, and $U({\bf{r}})$ the trapping potential. 
We show this calculation only for the the case of fermionic impurity, analogous arguments leading to neglecting the effect of the mean energy hold in the case of thermal bosonic impurities.

For degenerate impurities the mean energy recovers the value given by the low-temperature Sommerfeld expansion of the energy of an ideal trapped Fermi gas \cite{Pethick2008book}:
\beq
\bar\varepsilon  \overset{T\ll T_{\rm F_{\rm K}}}{\approx} \frac{3}{4} E_{\rm F_{\rm K}}(\mathcal{C})+ \frac{\pi^2}{2}k_{\rm B}\frac{T^2}{T_{\rm F_{\rm K}}(\mathcal{C})},
\eeq
where $E_{\rm F_{\rm K}}(\mathcal{C})$ and $T_{\rm F_{\rm K}}(\mathcal{C})$ are the Fermi energy and temperature of the impurities for a given concentration $\mathcal{C}$.

For thermal impurities we find a linear dependence on the impurity concentration $\mathcal{C}$~\cite{Pethick2008book}:
\beq
\label{HT_exp}
\bar\varepsilon  \overset{T\gg T_{\rm F}}{\approx} 3k_BT \left(1 + \frac{\gamma}{T^3}\mathcal{C}\right),
\eeq
where for a three dimensional harmonically trapped gas $\gamma = 0.056 N_{\rm Li} \left( \frac{\hbar \bar \omega_{\rm K}}{k_{\rm B}}  \right)^3$, with $\bar \omega_{\rm K}$ the geometrically averaged K trap frequency.

We numerically solve equation~\eqref{eq_scazza} by fixing the integration volume $\mathcal{V}$ to the full volume of the K cloud. The parameters are chosen to be close to typical experimental values: $T=250$ nK and atom number $N_{\rm Li}$ of Li such that $E_{\rm F_{\rm Li}} = 20$ kHz. In this way we can solve equation~\eqref{eq_scazza} for the experimental relevant values of the impurity concentration $\mathcal{C} = N_{\rm K}$/$N_{\rm Li} = 0-0.5$. 
\begin{figure}[!ht]
\centering
\includegraphics[width= 180mm]{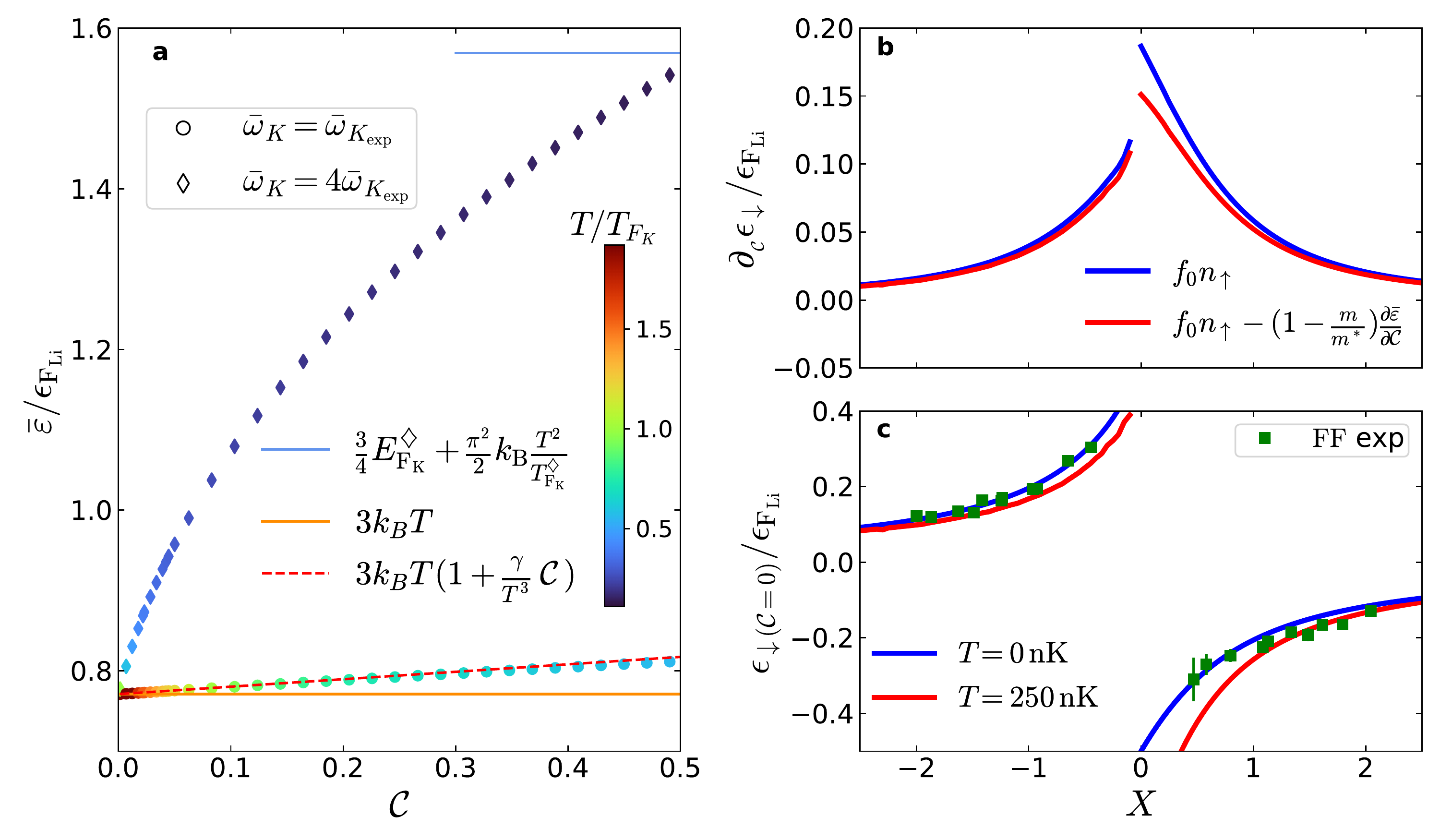} 
\caption{\label{fig:supinf_fig1} \textbf{Effects of the impurity mean energy.}
\textbf{a}: mean energy of the K impurities as a function of the concentration, obtained by numerically solving equation~\eqref{eq_scazza} with typical experimental parameters. Circles are obtained using the same value for the trap frequencies as in the experiment, diamonds using four fold increased ones. The color scale represent the effective temperature of the impurities for different concentrations. The orange and the blue lines represent the limit of a classical and an ideal Fermi gas at finite temperature, respectively. $E_{\rm F_{\rm K}}^{\diamondsuit}$ and $T_{\rm F_{\rm K}}^{\diamondsuit}$ are, respectively, the Fermi energy and the Fermi temperature relative to impurities with $\mathcal{C}=0.5$. The red dashed line represents the high temperature expansion.
\textbf{b}: variation of the polaron energy with the concentration, considering only the mediated polaron-polaron interaction (solid blue lines) and the mediated polaron-polaron interaction plus the contribution of the mean energy of the impurities (solid red lines). \textbf{c}: single impurity limit for $T=0$ (solid blue lines) and considering the effect of the impurity mean energy for $T=250$ nK (solid red lines). The green squares are the experimental data for the FF mixture, as in Fig.~3 of the main text. 
The quantities in all the panels are normalized by the Fermi energy of the Li Fermi sea}
\end{figure}

In order to observe the difference between a thermal and a degenerate gas, we perform the calculation of equation~\eqref{eq_scazza} using typical atom number and temperature as in the experiments, but for two different values of the K trap frequencies: the same value used in the experiment, $\bar \omega_{\rm K} = \bar \omega_{\rm K_{\rm exp}}$, and $\bar \omega_{\rm K} = 4\bar \omega_{\rm K_{\rm exp}}$. 
In this way we vary the Fermi energy of the impurities, and thus their effective temperature $T/T_{\rm F_{\rm K}}$, without varying atom number and absolute temperature.
The results are shown in Fig.~S\ref{fig:supinf_fig1}a.
We see that for our experimental parameters (circles) the impurities, which remains in the thermal regime, have a smaller increase of their mean energy with respect to the degenerate case (diamonds) by increasing their concentration, and can be described by the linear behaviour of equation~\eqref{HT_exp} (red dashed line).

Taking into account the effect of the mean energy, we can write the polaron energy as~\cite{Scazza2017rfp} 
\beq
\label{full_pol}
\epsilon_{\downarrow}=\epsilon_{\downarrow}^0+f_0 n_\uparrow \mathcal{C} - \left(1 -\frac{m_{\rm K}}{m_{\rm K}^*}\right)\bar \varepsilon ,
\eeq
where the minus sign in front of the last term comes for the fact that in our experiment the non interacting state has a lower energy than the interacting one and we used the static limit for the quasi-particle interaction $f_0$ defined in equation (3) of the main text.
In the main text, we identify the mediated polaron-polaron interaction coefficient with the value of the slope obtained from a linear fit of our experimental obtained polaron energy as a function of concentration.
In order to estimate the effect of neglecting the mean energy contribution in obtaining the mediated polaron-polaron interaction, we can compare the values of $\frac{\partial \epsilon_{\downarrow}}{\partial \mathcal{C}}$ with $\epsilon_{\downarrow}$ given by equation~\eqref{full_pol} neglecting and considering the last term, respectively, for thermal impurities. In particular we can use the high temperature expansion~\eqref{HT_exp} in order to obtain $\frac{\partial \bar \varepsilon}{\partial \mathcal{C}} = 3k_B\gamma/{T^2}$.
This comparison is reported in Fig.~S\ref{fig:supinf_fig1}b, from which we see that the effect of the mean energy is negligible. This is due to two main differences between our Li-K experiment and the Li-Li experiment of Ref.~\cite{Scazza2017rfp}: first, thanks to our heterogeneous mixture, we can reach degeneracy for Li, while keeping K thermal. This contributes to reducing the variation of the impurity mean energy when varying the impurity concentration.
Second, in our experiment $(1-\frac{m_{\rm K}}{m_{\rm K}^*} )$ is close to $0$ (e.g, $(1-\frac{m_{\rm K}}{m_{\rm K}^*} ) = 0.037 $ and $0.067$ at $X = -1$ and $X = 1$, respectively), reducing the overall contribution of the mean energy effect to the polaron energy.

We consider now the effect of neglecting the mean energy term on the determination of the polaron energy in the zero concentration limit. From equation~\eqref{HT_exp} we see that in this limit we recover the classical expression $\bar \varepsilon = 3k_BT$.
So we can compare our experimental results presented in Fig.~3 of the main text with $\epsilon_{\downarrow}^0- (1 -\frac{m_{\rm K}}{m_{\rm K}^*})3k_{\rm B}T$, as reported in Fig.~S\ref{fig:supinf_fig1}c. 
Comparing the theory predictions with and without the mean energy term (solid red and blu lines, respectively), we see that the main effect of the mean energy is given by a down shift of the single polaron energy. This shift is in line with what we observe in the experiment for $|X| > 1$. Interestingly, our experimental data seem to follow the zero temperature limit for stronger interaction ($|X| < 1$), where we see strong deviation from Landau's Fermi liquid theory.

\section{Many-body theory}

As explained in the Methods section, the mediated interaction between the polarons mediated by the Fermi sea can be obtained from the impurity self-energy. In the ladder approximation, this is given by
\begin{eqnarray} \label{eq:sigmaap}
\Sigma(k)=\frac{1}{\beta\mathcal{V}}\sum_{q}\mathcal{G}_{\uparrow}(q)\mathcal{T}(k+q)
\end{eqnarray}
where $\beta$ is the inverse temperature, $\mathcal{V}$ is the volume, and 
$\mathcal{G}_{\uparrow}(k)=1/[i\kappa_{\nu}-\xi_{\mathbf{k}\uparrow}]$ is the propagator of ideal majority fermions
with $\xi_{\mathbf{k}\uparrow}={\mathbf k}^2/2m_\uparrow-\varepsilon_{F}$. We have defined the four-momenta $k=(\mathbf{k},i\kappa_{\nu})$ and $q=(\mathbf{q},i\kappa'_{\nu})$ where $\kappa'_{\nu}$ is a fermionic Matsubara frequency while $\kappa_{\nu}$ is a bosonic/fermionic Matsubara frequency depending on the quantum statistics of the impurities. The many-body scattering matrix is given by 
\begin{gather} 
\mathcal{T}( k)=\mathcal{T}_{v}(k)\left[1-\mathcal{T}_{v}(k)\Pi(k)\right]^{-1}. 
\label{eq:Tmatrix}
\end{gather} 
where
\begin{gather}
\mathcal{T}_{v}(k)=\frac{2\pi}{m_{r}}\left[\frac{1}{a}+R^{*}k_{r}^{2}(k)\right]^{-1},
\end{gather}
with
\begin{gather}
k_{r}(k)=\sqrt{i\kappa_{\nu}-\frac{k^{2}}{2M}+\varepsilon_{F}}
\end{gather}
is the scattering matrix in the vacuum, including finite range effects of the interaction~\cite{Fritsche2021sab}.
Here,  $a$ is the scattering length, $R^{*}$ is a range parameter, $M=m_{\downarrow}+m_{\uparrow}$, $m_r=m_{\downarrow}m_{\uparrow}/(m_{\downarrow}+m_{\uparrow})$, $m_{\downarrow}$ and $m_{\uparrow}$ the total, reduced, impurity and medium particles masses, respectively. The pair propagator for an $\uparrow$ and a $\downarrow$ atom is 
\begin{gather} \label{eq:ps}
\Pi(q)=\int \frac{d^{3}\mathbf{p}}{(2\pi)^{3}}\left[\frac{1 \pm \,\text{f}_{\downarrow}(\xi_{\mathbf{p}+\mathbf{q}\downarrow})-\text{f}_{\uparrow}(\xi_{-\mathbf{p}\uparrow}) }{i\kappa_{\nu}-\xi_{\mathbf{p}+\mathbf{q}\downarrow}-\xi_{-\mathbf{p}\uparrow}}+\frac{2m_r}{\mathbf{p}^{2}}\right],
\end{gather}
with $\xi_{\mathbf{p}\downarrow}=\epsilon_{\mathbf{p}\downarrow}-\mu_{\downarrow}$ and $\mu_{\downarrow}$ the chemical potential of the impurities. The upper and lower sign corresponds to bosonic and fermionic impurities, respectively, with a distribution function $\text{f}_{\downarrow}(\xi)=1/[\exp(\beta\xi)\mp 1]$. For the fermionic majority atoms, the distribution function is $\text{f}_{\uparrow}(\xi)=1/[\exp(\beta\xi)+1]$. Performing the Matsubara sum in equation~\eqref{eq:sigmaap} gives~\cite{Bastarrachea2021aar,Bastarrachea2021pia}
\begin{gather} \label{eq:ssap}
\Sigma(k)=\Sigma_{\text{p}}(k)+\Sigma_{\text{b.c.}}(k)+\Sigma_{\text{m}}(k),
\end{gather}
where
\begin{gather} \label{eq:ssap1}
\Sigma_{\text{p}}(k)= \int\frac{d^{3}\mathbf{q}}{(2\pi)^{3}}\text{f}_{\uparrow}(\xi_{\mathbf{q}\uparrow})\mathcal{T}(\mathbf{k}+\mathbf{q},i\kappa_{\nu}+\xi_{\mathbf{q}\uparrow}), \\
\label{eq:ssap2}
\Sigma_{\text{b.c.}}(k)= \pm\int\frac{d^{3}\mathbf{q}}{(2\pi)^{3}}\int_{-\infty}^{\infty}\,\frac{d\omega'}{\pi}\,\frac{\text{f}_{\text{m}}(\omega')\mbox{Im}\mathcal{T}(\mathbf{k}+\mathbf{q},\omega'+i0^{+})}{i\kappa_{\nu}-\omega'+\xi_{\mathbf{q}\uparrow}}, 
\end{gather}
and
\begin{gather}
\label{eq:ssap3}
\Sigma_{\text{m}}(k)= \mp\int\frac{d^{3}\mathbf{q}}{(2\pi)^{3}}\frac{\text{f}_{\text{m}}(\omega_{\text{m}\mathbf{k}+\mathbf{q}})\text{Z}_{\text{m}\mathbf{k}+\mathbf{q}}}{i\kappa_{\nu}-\omega_{\mathbf{k}+\mathbf{q}\text{m}}+\xi_{\mathbf{q}\uparrow}}.
\end{gather}
The first term $\Sigma_{\text{p}}$ is non-zero even when there is only one impurity (zero concentration) and thus determines the energy of a single polaron. The last two terms are present only for non-zero impurity concentrations. They are crucial to obtain a reliable expression for the mediated interaction. The second term $\Sigma_{\text{b.c.}}$ comes from the branch cut of $\mathcal{T}$-matrix corresponding to the impurity-atom continuum. The last term $\Sigma_{\text{m}}$ is present only when $\mathcal{T}(\mathbf{k}+\mathbf{q},\omega)$ has a pole at $\omega=\omega_{\mathbf{k}+\mathbf{q}\text{m}}$, which corresponds to a Feshbach molecule consisting of one impurity and one majority atom with total momentum $\mathbf{k}+\mathbf{q}$. These molecules emerge close to resonance, and $\text{Z}_{\text{m}\mathbf{k}}=1/\left.\partial_{\omega}\text{Re}\left[\mathcal{T}(\mathbf{k},\omega)\right]^{-1}\right|_{\omega_{\mathbf{k}\text{m}}}$ is the corresponding residue of the $\mathcal{T}$-matrix. 
Their distribution function $\text{f}_{\text{m}}(\xi)=1/[\exp(\beta\xi)\pm 1]$ is fermionic or bosonic depending on whether the impurity is bosonic or fermionic.

To get the mediated interaction between the polarons, we analytically continue the self-energy from Matsubara to real frequencies. Subsequently, we expand it to linear order in the impurity and molecule populations. A tedious but straightforward calculation gives  $\Sigma(k)=\Sigma_0(k)+\delta\Sigma(k)$ where $\Sigma_0(k)$ is the self-energy for a single impurity, i.e.\ the value of $\Sigma_p(k)$ for vanishing impurity concentration, and 
\begin{align} \label{eq:dsap} 
\delta\Sigma(k)=&
\pm\int\frac{d^{3}\mathbf{p}}{(2\pi)^{3}} \text{f}_{\downarrow}(\xi_{\mathbf{p}\downarrow}) \int\frac{d^{3}\mathbf{q}}{(2\pi)^{3}}  \frac{\text{f}_{\uparrow}(\xi_{\mathbf{q}\uparrow})\mathcal{T}_{0}^{2}(\mathbf{k}+\mathbf{q},\omega+\xi_{\mathbf{q}\uparrow})-\text{f}_{\uparrow}(\xi_{\mathbf{k}+\mathbf{q}-\mathbf{p}\uparrow})|\mathcal{T}_{0}(\mathbf{k}+\mathbf{q},\omega+\xi_{\mathbf{p}\uparrow}+\xi_{\mathbf{k}+\mathbf{q}-\mathbf{p}\downarrow})|^{2}}{\omega+i0_+-\xi_{\mathbf{p}\downarrow}+\xi_{\mathbf{q}\uparrow}-\xi_{\mathbf{k}+\mathbf{q}-\mathbf{p}\uparrow}}\nonumber\\
&\mp\int\frac{d^{3}\mathbf{q}}{(2\pi)^{3}}\frac{\text{f}_{\text{m}}(\omega_{\mathbf{k}+\mathbf{q}\text{m}})\text{Z}_{\text{m}\mathbf{k}+\mathbf{q}}}{\omega+i0_+-\omega_{\mathbf{k}+\mathbf{q}\text{m}}+\xi_{\mathbf{q}\uparrow}}.
\end{align}
Here, $\mathcal{T}_{0}$ is the many-body scattering matrix in the zero impurity concentration limit. As described in the Methods section, one can now extract the mediated interaction between polarons from the first term in equation \eqref{eq:dsap}, and between polarons and molecules from the second term in equation~\eqref{eq:dsap}. The first term gives the linear dependence of the impurity self-energy and therefore, the polaron energy on the impurity population $\text{f}_{\downarrow}(\xi_{\mathbf{p}\downarrow})$. For weak impurity-atom interaction, one has $\mathcal{T}(k)\simeq 2\pi a/m_{r}$ and there are no Feshbach molecules, so the second term in equation~\eqref{eq:dsap} is absent, and one obtains
\begin{align}
\delta\Sigma(k)\simeq & \, \pm\left(\frac{2\pi a}{m_{r}}\right)^{2}\int\!\frac{d^{3}\mathbf{p}}{(2\pi)^{3}} \text{f}_{\downarrow}(\xi_{\mathbf{p}\downarrow}) \int\!\frac{d^{3}\mathbf{q}}{(2\pi)^{3}}  \frac{\text{f}_{\uparrow}(\xi_{\mathbf{q}\uparrow})-\text{f}_{\uparrow}(\xi_{\mathbf{k}+\mathbf{q}-\mathbf{p}\uparrow})}{\omega+i0_+-\xi_{\downarrow\mathbf{p}}+\xi_{\uparrow\mathbf{q}}-\xi_{\uparrow\mathbf{k}+\mathbf{q}-\mathbf{p}}}\nonumber\\\label{RKKY}
=& \, \pm\left(\frac{2\pi a}{m_{r}}\right)^{2}\int\!\frac{d^{3}\mathbf{p}}{(2\pi)^{3}} \text{f}_{\downarrow}(\xi_{\mathbf{p}\downarrow})\,\chi(\mathbf{k}-\mathbf{p},\omega-\xi_{\mathbf{p}\downarrow}),
\end{align}
where $\chi(\mathbf{k},\omega)$ is the Lindhard function~\cite{Giuliani2005}. Equation~\eqref{RKKY} is the well-known second-order RKKY result for the mediated interaction mediated by particle-hole excitations in a Fermi gas~\cite{Ruderman1954,Kasuya1956,Yosida1957}. Taking the limit of zero energy and momenta  yields the weak-coupling limit of equation (3) in the 
main manuscript (since $\Delta N\propto a$ for small $k_F|a|$). 
 
The second term in equation~\eqref{eq:dsap} gives the energy shift due to a population of Feshbach molecules,  thus corresponding to a molecule-polaron interaction. This interaction can be interpreted as coming from a Bose stimulation/Fermi blocking of a second-order shift of the polaron energy due to the coupling to Feshbach molecules~\cite{Bastarrachea2021aar,Bastarrachea2021pia}. 
When modeling the experimental data, we use equation~\eqref{eq:dsap} for a zero momentum polaron with the impurity and molecule concentrations determined via their chemical potentials.

%
